# The hydrostatic equilibrium and Tsallis' equilibrium for self-gravitating systems

Du Jiulin*

*Department of Physics, School of Science, Tianjin University, Tianjin 300072, China*

**Abstract**

Self-gravitating systems are generally thought to behavior non-extensively due to the long-range nature of gravitational forces. We obtain a relation between the nonextensive parameter $q$ of Tsallis statistics, the temperature gradient and the gravitational potential based on the equation of hydrostatic equilibrium of self-gravitating systems. It is suggested that the nonextensive parameter in Tsallis statistics has a clear physical meaning with regard to the non-isothermal nature of the systems with long-range interactions and Tsallis' equilibrium distribution for the self-gravitating systems describes the property of hydrostatic equilibrium of the systems.

------------------

* Email: jiulindu@yahoo.com.cn or jldu@tju.edu.cn

The hydrostatic equilibrium is one of fundamental properties (or one of the basic assumptions) of self-gravitating systems. It has been long understood that the self-gravitating systems that are at stable state satisfy the equation of hydrostatic equilibrium, the general form of which can be written as

$$\nabla P = -mn\nabla \varphi(\mathbf{r}) \tag{1}$$

where $P$ is the pressure, $m$ is the mass of particle, $n$ is the number density of particles, and $\varphi$ is the gravitational potential determined by Poisson equation,

$$\nabla^2 \varphi(\mathbf{r}) = 4\pi Gmn \tag{2}$$

where $G$ is the gravitational constant. In the framework of Boltzmann-Gibbs (B-G) statistical mechanics, the structure and stability of self-gravitating systems at statistical equilibrium are usually analyzed in terms of the maximization of a thermodynamic potential [1]. This thermodynamic approach leads to isothermal configurations that have been studied for long time in the context of stellar structure and galactic structure with equation of state of idea gas taken in the form of $P = nkT$ and Maxwell-Boltzmann (M-B) equilibrium distribution expressed by

$$f(\mathbf{r},\mathbf{v}) = \left(\frac{m}{2\pi kT}\right)^{\frac{3}{2}} n(\mathbf{r}) \exp\left(-\frac{m\mathbf{v}^2}{2kT}\right) \tag{3}$$

where temperature $T$ is constant. The number density of particles is given by

$$n(\mathbf{r}) = n_0 \exp[-\frac{m}{kT}(\varphi - \varphi_0)] \tag{4}$$

where $n_0$ and $\varphi_0$ are, respectively, the number density and the gravitational potential at $r = 0$. Additionally, Eq.(3) and Eq.(4) can be obtained by following the standard line of using Boltzmann equation and $H$ theorem [2], which leads to the result known to all that the temperature gradient is zero and therefore the system is a thermal equilibrium state. The density distribution Eq.(4) can also be determined more directly by combining the equation of hydrostatic equilibrium, Eq.(1), with the pressure above in the equation of state of idea gas.

Self-gravitating systems have been generally thought to behavior non-extensively due to the long-range nature of gravitational forces. However, almost all the systems treated in statistical mechanics with B-G statistics have usually been extensive; this property holds for systems with short-range interparticle forces. When we deal with systems with long-rang interparticle forces such as Newtonian gravitational forces and Coulomb electric forces, where nonextensivity holds, B-G statistics may need to be generalized for the statistical description of such systems. The nonextensive generalization of B-G statistical mechanics known as "Tsallis statistics" has focused significant attention in recent years [3]. Such a generalization was done by constructing a new form of entropy, $S_q$, with the nonextensive parameter $q$ different from unity [4] in the form

$$S_q = \frac{k}{1-q}(\sum_i p_i^q - 1) \qquad (6)$$

where $k$ is Boltzmann constant, $p_i$ is probability that the system under consideration is in its $i$th configuration such that $\sum_i p_i = 1$, and $q$ is a positive parameter whose deviation from unity is considered for describing the degree of nonextensivity of the system. The Boltzmann entropy $S_B$ is recovered from $S_q$ only if $q=1$. In this way, Tsallis statistics gives for all $q \neq 1$ a power law distribution, while the B-G exponential distribution is obtained only for $q=1$. This new theory has provided a convenient frame for the thermo-statistical analyses of many astrophysical systems and processes [5], such as Jeans criterion for self-gravitating systems [6-8], stellar polytropes [9-11], galaxy clusters [12], the nonequilibrium dynamical evolution of stellar systems [13,14], dark matter distribution [15], the negative specific heat [16,17] and the solar neutrino problem [18,19], etc.

Recently, M-B equilibrium distribution has been generalized based on the generalized Boltzmann equation in the framework of Tsallis' statistics [20]. Tsallis' equilibrium distribution (or the generalized M-B distribution) for the self-gravitating systems is also analyzed [21], which leads to a power law expression as

$$f_q(\mathbf{r},\mathbf{v}) = n(\mathbf{r})B_q \left(\frac{m}{2\pi kT(\mathbf{r})}\right)^{\frac{3}{2}} \left[1-(1-q)\frac{m\mathbf{v}^2}{2kT(\mathbf{r})}\right]^{\frac{1}{1-q}} \quad (7)$$

where $B_q$ is a $q$-dependent normalization constant and the density distribution is temperature-dependent,

$$n(\mathbf{r}) = n_0 \left(\frac{T(\mathbf{r})}{T_0}\right)^{\frac{3}{2}} \exp\left[-\frac{m}{k}\left(\int \frac{\nabla \varphi(\mathbf{r})}{T(\mathbf{r})} \cdot d\mathbf{r} - \frac{\varphi_0}{T_0}\right)\right] \quad (8)$$

where $T_0$ is temperature at $r=0$. The nonextensive parameter $q$ for the systems is derived with relation to the gravitational potential and the temperature gradient by the following equation [21]

$$k\nabla T + (1-q)m\nabla\varphi = 0 \quad (9)$$

It is therefore presented a physical meaning of $q$ with regard to the nature of non-isothermal configurations of self-gravitating systems. The nonextensive parameter $q$ is not one if and only if the temperature gradient is not zero. M-B equilibrium distribution is recovered from Eq.(7) perfectly when we let $q=1$.

In the framework of Tsallis' statistics, the equation of state of idea gas having been used for long time in astrophysics is modified due to the nonextensive effect of long-range interactions of gravitational forces [7, 8]. We now consider the equation of hydrostatic equilibrium for self-gravitating systems in Tsallis' statistics. We first discuss the nonextensive equation of state of the system. In Tsallis' equilibrium distribution Eq.(7), there is a thermal cutoff on the maximum value allowed for the velocity of a particle for $q<1$, $v_{\max} = \sqrt{2kT/m(1-q)}$, whereas there is no thermal cutoff for $q>1$, $v_{\max} \to \infty$. Then the mean value of square velocity $<v^2>$ is expressed by the integral

$$<v^2> = \int_0^{v_{\max}} v^2 f_q(\mathbf{r},\mathbf{v})\, d^3v \quad (10)$$

Substituting Eq.(7) into Eq.(10), we can evaluate this mean value for $q<1$ by making the change of variables, $\mathbf{u} = [(1-q)m/2kT]^{1/2}\mathbf{v}$. So we have

$$<v^2> = \left[(1-q)\frac{m}{2kT}\right]^{-1} \int_0^1 u^4(1-u^2)^{\frac{1}{1-q}} du \bigg/ \int_0^1 u^2(1-u^2)^{\frac{1}{1-q}} du$$

$$= \frac{6kT}{m(7-5q)}. \tag{11}$$

Obviously, the standard mean value of square velocity, $3kT/m$, in B-G statistics is recovered from Eq.(11) if we take $q =1$. It is easy to prove that the result in Eq.(11) still holds true for $1 \le q < 7/5$. But for $q \ge 7/5$, the mean value of $v^2$ diverges. Thus, the nonextensive equation of state of the self-gravitating system is obtained as

$$P = \frac{1}{3}mn<v^2> = \frac{2}{7-5q}nkT, \quad 0 < q < \frac{7}{5}. \tag{12}$$

The standard form of the equation of state of idea gas is recovered by taking $q =1$. Using this new equation of state instead of that one of idea gas, we can write the equation of hydrostatic equilibrium, Eq.(1), as

$$\frac{2}{7-5q}k(n\nabla T + T\nabla n) = -mn\nabla\varphi \tag{13}$$

where the density gradient $\nabla n$ can be determined from Eq.(8), which is expressed with the gravitational potential and the temperature gradient by the equation

$$\frac{\nabla n}{n} = \frac{3\nabla T}{2T} - \frac{m\nabla\varphi}{kT} \tag{14}$$

Substituting Eq.(14) into Eq.(13), we can obtain the relation between the nonextensive parameter $q$, the gravitational potential and the temperature gradient,

$$k\nabla T + (1-q)m\nabla\varphi = 0 \tag{15}$$

It is clear that this relation is the same as that one in Eq.(9), which was determined quite generally by the generalized Boltzmann equation, $q$-$H$ theorem and Tsallis' equilibrium distribution [21, 22]. Thus, again we obtain Eq.(9) from the equation of hydrostatic equilibrium for self-gravitating systems. This excellent agreement between the results obtained in tow different ways suggests that Tsallis' equilibrium distribution for self-gravitating systems describes the property of hydrostatic equilibrium of the systems.

We are usually to say that Tsallis statistics for $q \neq 1$ can be used to describe the systems with long-range interactions. However, we do not know why it can do so. In other words, we do not know whether the parameter $q \neq 1$ must be related to a long-range potential. In our present work, Eq.(9) or Eq.(15) establish a close relation between the parameter $q \neq 1$, the gravitational potential $\varphi$ and the temperature gradient.

But, from the mathematical point of view alone, it is not necessary for $\varphi$ to be a long-range potential and it can be any one. Actually, from the physical point of view, that the nonextensive parameter $q \neq 1$ is related to the long-range nature of gravitational forces in the self-gravitating system can be understood reasonably by Eq.(15) or Eq.(9). If the potential $\varphi$ is short-range interactions, then each element in the system is free from the boundary of the system. Such a system is the physical "large" one. The physical "large" system is always to limit to thermal equilibrium because each particle of the system is free, the temperature gradient is zero and $q$ is unity. But, if the potential $\varphi$ is long-range interactions, then each element in the system is feeling the boundary of the system and the system is to behavior physical "small" (here not real small in space scale but "small" in the physical meaning). The physical "small" system cannot reach to thermal equilibrium *automatically* because each particle in the system is not free and it is always under control of the long-range potential. So, the temperature gradient cannot be zero and $q$ is not unity. This may be the reason why the self-gravitating system is always at the non-isothermal state if the convective mixing is not taking place.

In summary, the nonextensive parameter in Tsallis statistics has a clear physical meaning with regard to the non-isothermal nature of the systems with long-range interactions. Tsallis' equilibrium distribution for self-gravitating systems describes the fundamental property of hydrostatic equilibrium of the systems.

**Acknowledgments**

I would like to thank S.Abe, C.Tsallis, P.Quarati, H.J.Haubold and A.Q.Wang for helpful discussions during the twelfth UN/ESA workshop on Basic Space Science at Beijing. This work is supported by the project of "985" Program of TJU of China.

**References**

[1] S.Chandrasekhar, *An introduction to the theory of stellar structure* (Dover, 1942); J.Binney, S.Tremaine, *Galactic dynamics* (Princeton Series in Astrophysics, 1978).

[2] S. Chapman, T. G. Cowling, *The Mathematical Theory of Nonuniform Gases* (Third Edition, Cambridge University Press, 1970); Wang Zhuxi, *An Introduction to Statistical Physics* (The People's Education Press, Beijing, 1965).

[3] S.Abe, A.K.Rajagopal, Science, **300**(2003)249; A. Plastino, Science **300**(2003)250; V. Latora, A. Rapisarda and A. Robledo, Science **300**(2003)250; R.V. Chamberlin, Science **298**(2002)1172.

[4] C.Tsallis, J.Stat.Phys. **52**(1988)479.

[5] C.Tsallis, D.Prato, *Nonextensive statistical mechanics*: *Some links with astronomical phenomena*, in the Proceedings of the Xith United Nationa / European Space Agency Workshop on Basic Space Sciences, Office for Outer Space affairs / United Nations (Cordoba, 9-13 Sept. 2002), edited by H.J.Haubold.

[6] J.A.S.Lima, R.Silva and J.Santos, Astron.Astrophys. **396**(2002)309.

[7] J.L.Du, Phys.Lett.A **320**(2004)347.

[8] J.L.Du, Physica A **335**(2004)107.

[9] A.Taruya and M.Sakagami, Physica A **307**(2002)185; M.Sakagami and A.Taruya, Physica A **340**(2004)444.

[10] A.Plastino and A.R.Plastino, Phys.Lett.A **174**(1993)384.

[11] R.Silva and J.S.Alcaniz, Physica A **341**(2004)208.

[12] C.A.Wuensche, A.L.B.Ribeiro, F.M.Ramos and R.R.rosa, Physica A **334**(2004)743.

[13] A.Taruya and M.Sakagami, Phys.Rev.Lett. **90**(2003)181101.

[14] A.Taruya and M.Sakagami, Physica A **340**(2004)453.

[15] S.H.Hansen, D.Egli, L. Hollenstein and C.salzmann, astro-ph/ 0407111, to appear in MNRAS (2004).

[16] S.Abe, Phys.Lett.A **263**(1999)424.

[17] R.Silva and J.S.Alcaniz, Phys.Lett.A **313**(2003)393.

[18] G.Kaniadakis, A.Lavagno and P.Quarati, Phys.Lett.B **369**(1996)308; A.Lavagno and P.Quarati, Chaos,Solitons, Fractals **13**(2002)569.

[19] M.Coraddu, M.Lissia, G.Mezzorani and P.Quarati, Physica A **326**(2003)473.

[20] J.A.S.Lima, R.Silva, A.R.Plastino, Phys.Rev.Lett. **86**(2001)2938.

[21] J.L.Du, Europhys.Lett. **67**(2004)893.

[22] J.L.Du, Phys.Lett.A **329**(2004)262.